\documentclass[12pt]{article}
\title{\begin{flushright}
{\normalsize McGill/99-37\\NUC-MINN-99/16-T\\}
\end{flushright}
{\bf Coherence Time Effects on $J/\psi$ Production and Suppression in
Relativistic Heavy Ion Collisions}}
\author{{\bf Charles Gale}\thanks{gale@physics.mcgill.ca}\vspace*{0.1in} \\
 {\it Physics Department}\\
 {\it McGill University}\\ \vspace*{0.2in}
 {\it Montreal, Quebec H3A 2T8, Canada}\\ \vspace*{0.1in}
{\bf Sangyong Jeon}\thanks{jeon@nta2.lbl.gov} \\
 {\it Nuclear Science Division}\\
 {\it Lawrence Berkeley National Laboratory} \\ \vspace*{0.2in}
 {\it Berkeley, CA 94720, USA}\\ \vspace*{0.1in}
{\bf Joseph Kapusta}\thanks{kapusta@physics.spa.umn.edu}\\
  {\it School of Physics and Astronomy}\\
  {\it University of Minnesota}\\ {\it Minneapolis, MN 55455, USA}}

\date{}

\usepackage[dvips]{graphicx}

\begin{document}
\maketitle

\begin{abstract}
Using a coherence time extracted from high precision proton-nucleus
Drell-Yan measurements and a nuclear absorption cross section
extracted from $pA$ charmonium production experiments, we study
$J/\psi$ production and absorption in nucleus-nucleus collisions. We
find that coherence time effects are large enough to affect the measured 
$J/\psi$-to-Drell-Yan ratio. The S+U data at 200A GeV/c measured by NA38 are reproduced
quantitatively without the introduction of any new parameters. However, when
compared with recent NA50 measurements for Pb+Pb at 158A GeV/c, 
the data is not reproduced in trend or in magnitude.\\

\end{abstract}

\vspace*{1cm}


Ultrarelativistic heavy ion collisions offer the tantalizing
possibility of forming and studying a new form of matter predicted by 
QCD: the quark-gluon plasma. A vigorous
experimental program has existed at the CERN SPS for more than ten
years.  RHIC now signals the dawn of a new era in heavy ion physics 
at Brookhaven National Laboratory. Several experimental signals have 
been put forward as candidates for QCD plasma signatures \cite{muller}.
  Of those, the most famous is probably that of $J/\psi$ suppression
in nucleus-nucleus collisions. The theoretical and 
experimental activity that have followed this seminal suggestion have been
considerable as the disappearance of the $J/\psi$ can directly be
linked to deconfinement and Debye screening in the plasma \cite{ms86}.  
The interested reader will find a recent
snapshot of the state of this field in Ref. \cite{qm97}. 

Before an experimentally observed $J/\psi$ suppression pattern is
interpreted as an unambiguous signal of the existence of a quark-gluon
plasma, it is imperative to rule out all competing explanations of purely
hadronic origin. Moreover, the hadronic scenarios considered should
incorporate elements of physics that are known to be relevant at the energy
scale under consideration. It is one such line of thought that we follow in this
paper. 
We study charmonium production in
relativistic heavy ion collisions along with the appropriate
background, Drell-Yan pair production. 
Our paper is organized as follows: First we recall the main
features of a 
model that is successful in explaining high
precision Drell-Yan data measured in proton-nucleus collisions. 
Those data enable one to extract a formation time characteristic of
the emission of soft hadrons, essentially pions. Next we recall the
application of this model to the production of $J/\psi$ in $pA$ collisions.
From those measurements we have extracted a cross section for 
$J/\psi$ absorption on the nucleon. With this formulation we make
parameter-free calculations for nucleus-nucleus collisions and compare them
with experimental data. 


In almost all considerations involving heavy ion collisions at any
energy, the issues of dynamics and elementary processes
remain intimately connected and inseparable. In view of this, a successful 
modeling of nuclear collisions is a necessary prerequisite to a
deeper exploration of the fine points of the nuclear dynamics. To simulate the
heavy ion collision we prefer to work with hadronic variables
rather than partonic ones, and  make a straightforward linear 
extrapolation from proton-proton
scattering.  This extrapolation, referred to as LEXUS, was detailed
and applied to nucleus-nucleus collisions at beam energies of several
hundred GeV per nucleon in Ref. \cite{lexus}.  Briefly, the
inclusive distribution in rapidity $y$ of the beam proton in an
elementary proton-nucleon collision is parameterized rather well by
\begin{equation}
W_1(y) = \lambda \frac{\cosh y}{\sinh y_0}
+ (1-\lambda)\delta(y_0-y) \, ,
\label{eq1}
\end{equation}
where $y_0$ is the beam rapidity in the lab frame.
The parameter $\lambda$ has the value 0.6 independent of beam
energy, at least in the range in which it has been measured, which
is $12-400$ GeV \cite{lambda}.  It may be interpreted as the fraction of all
collisions which are neither diffractive nor elastic.
As a nucleon cascades through the nucleus its
energy is degraded. An underlying assumption in this model is that of straight
line trajectories.   In the case of a nucleus-nucleus collision, one obtains
the single-particle rapidity distribution of the $m$'th projectile 
nucleon after a collision with the $n$'th target nucleon through 
the solution of an evolution equation \cite{lexus}:
\begin{eqnarray}
W^P_{m, n} ( y ) = \int d y_P dy_T W^P_{m, n-1} (y_P ) W^P_{m - 1, n} (y_0 -
y_T) \, Q ( y - y_T , y_P - y_T , y - y_P )\ .
\label{evol}
\end{eqnarray}
In the above, the kernel is 
\begin{eqnarray}
Q (s, t, u) = \lambda \frac{\cosh s}{\sinh t} + (1 - \lambda ) \, \delta ( u )\ ,
\end{eqnarray}
originating from Eq. (\ref{eq1}). 
Equation (\ref{evol}) is a Boltzmann-like equation that is solved numerically. 
This rapidity distribution then gets folded with impact parameter over the density
distributions of the projectile and target nuclei, using a method 
described in detail in Ref. \cite{lexus}. 

Recently we have extracted the quantum 
coherence time needed to reproduce Drell-Yan pair production data in
$pA$ collisions. This can also be formulated in terms of the 
Landau-Pomeranchuk-Migdal 
effect \cite{lpm}.  We briefly recall the procedure followed. We 
began by computing the Drell-Yan yield at leading-order (LO) with the GRV
structure functions \cite{grv} with a K factor. Those structure functions
reflect a flavor-asymmetric Dirac sea.  Adopting a fixed K factor of 2.1, 
we compared the results to  pp Drell-Yan data 
at 800 GeV/c \cite{ppdy} and found the
agreement to be excellent for all measured values of $x_F$.  

Turning then to the case of proton-nucleus
collisions as measured by the E772 collaboration \cite{e772}, we 
deduced \cite{prl} that the formation (or coherence) time
needed to fit the measured $\sigma_{pA}^{DY}/\sigma_{pD}^{DY}$ ratios at different
values of $x_F$ is $\tau_0$~=~0.4~$\pm$~0.1~fm/c, in the frame of
the colliding nucleons. Practically, this coherence time can be related to 
an initial state
energy loss for some of the Drell-Yan producing collisions \cite{kkg} as 
follows.  In
LEXUS, we assumed that the energy available to produce a Drell-Yan pair was
that which the proton had after $n$ previous collisions. 
In order to reproduce the 800 GeV/c E772 data, we needed $n = 5 \pm 1$. 
The $n$ collisions correspond to a path length 
of $n / \sigma^{tot}_{NN} \rho $ in the target nucleus rest frame, 
where $\sigma^{tot}_{NN}$ is a total cross section of 40 mb, and
$\rho$ is a nuclear matter density of 0.155 nucleons/fm$^3$. 
Lorentz-transforming to the
nucleon-nucleon  center of mass, one obtains the value of the proper coherence 
time quoted above. 
In this language, a traditional Glauber-type 
model (with no energy
loss) would have $n = \infty$. Fixing $\tau_0$, the 
range in $n$ appropriate
for the energies being considered in this work (nucleon momenta of 158 GeV/c and 
200 GeV/c) is found to be $2 \le n \le  3$.

We then investigated $J/\psi$ production in $pA$ collisions
\cite{pl,na38}. The additional 
input needed there was the cross section for producing $J/\psi$ in elementary
nucleon-nucleon interactions. We used a parametrization that
follows from a tabulation of data by Louren\c{c}o \cite{na38}:  
\begin{eqnarray}
B \sigma_{N N \rightarrow J/\psi} ( x_F > 0 )\ = \ 37\, (1 - m_{J/\psi} /
\sqrt{s}\, )^{12}\
{\rm nb}\ .
\label{sigparm}
\end{eqnarray}
Here, $\sqrt{s}$
in the center of mass energy of the nucleon pair and $B$ is the branching
ratio into a muon pair. Using the functional $x_F$ dependence of the 
differential cross
section as measured by E789 \cite{e789}, one can write 
a normalized differential cross section to use as an input in LEXUS:  
\begin{eqnarray}
\frac{d\sigma_{NN \rightarrow J/\psi}}{dx_F} = 6 \sigma_{NN \rightarrow
J/\psi}(x_F > 0) (1-|x_F|)^5 \, .
\end{eqnarray}
From our
analysis \cite{pl}, we have extracted a $J/\psi$ absorption cross section in
nuclear matter of 3.6 mb. It is worthwhile to note that this value is in 
numerical agreement with the same quantity 
deduced from experiments of $J/\psi$ photoproduction on nuclei \cite{photo}.
It is smaller than that used in other phenomenological heavy ion applications
\cite{kln97}.  
The parameters in this model are thus completely determined by proton-nucleus
data. 


We now turn to recent experiments on the production of the
$J/\psi$ in S+U \cite{na38_2} collisions at 200A GeV/c and in Pb+Pb \cite{na50} 
collisions at 158A GeV/c, at the CERN SPS. Because the $J/\psi$ is measured
through its decay into dimuons, the production cross section has
traditionally been divided by the natural background in the
appropriate invariant mass region: that of Drell-Yan pairs. However, since the
absolute cross section measurements are now available, we will first 
verify the predictions of our model there.
Including the appropriate respective detector acceptance, the results 
for absolute cross sections are show in Table \ref{table1}. 

\begin{figure}
\vspace{-1cm}
\begin{center}
\includegraphics[angle=0, width=8cm]{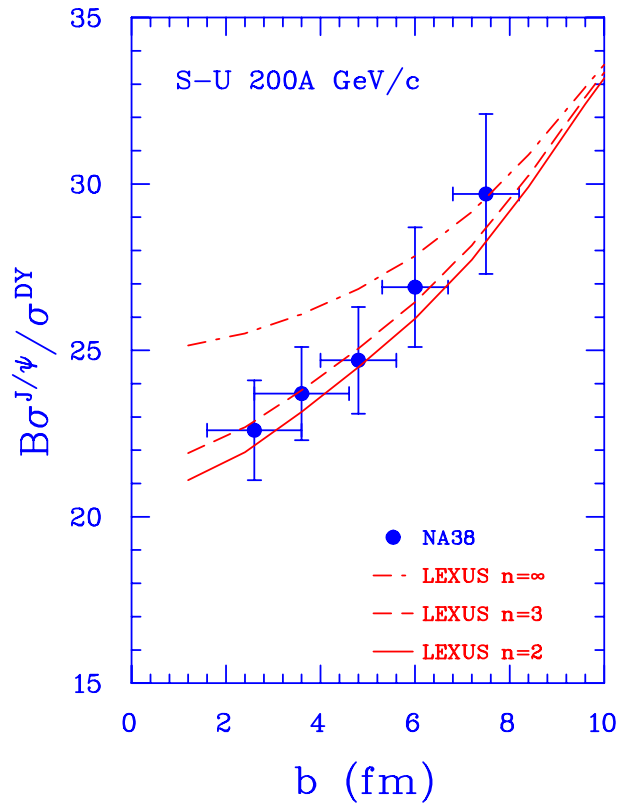}
\end{center}
\caption{Ratio of $J/\psi$ to Drell-Yan production cross sections. $B$ is
the branching ratio into a muon pair. The data is from Ref.
\protect\cite{na38_2}.}
\label{fig1}
\end{figure}

Consider the system S+U at 200A GeV/c. The coherence
time arguments made earlier in this paper suggest that the values $n = 2$
and $n = 3$ should bracket the experimental data. We observe that indeed this is
so, both for the measured Drell-Yan and $J/\psi$ absolute cross sections.
One can go further and plot $B \sigma^{J/\psi}/\sigma^{DY}$ against collision
centrality. One needs a model to map the impact parameter bins that enter
as input in our dynamical model into bins of measured transverse energy. The
experimental collaboration has in fact provided
the impact parameter range that corresponds to a measured $J/\psi$-Drell-Yan ratio
\cite{na38_2}. Comparison to the experimental data
is shown in Fig. \ref{fig1}. One can see that our results are consistent
with the data within experimental uncertainties.  Again, we emphasize that 
no new parameters were introduced. It is
also worthwhile to note that the numerator and denominator of the
plotted ratio have been calculated from ``first principles'', the meaning of
which is clear in the context of this paper: LO Drell-Yan 
calculations and a parametrization of the differential $J/\psi$ production 
in nucleon-nucleon collisions. 
The absolute cross
sections calculated with no energy loss (or infinite coherence time) 
fail to reproduce the experimental results. This is also the case for their
ratio.

\begin{figure}
\begin{center}
\includegraphics[angle=0,width=8cm]{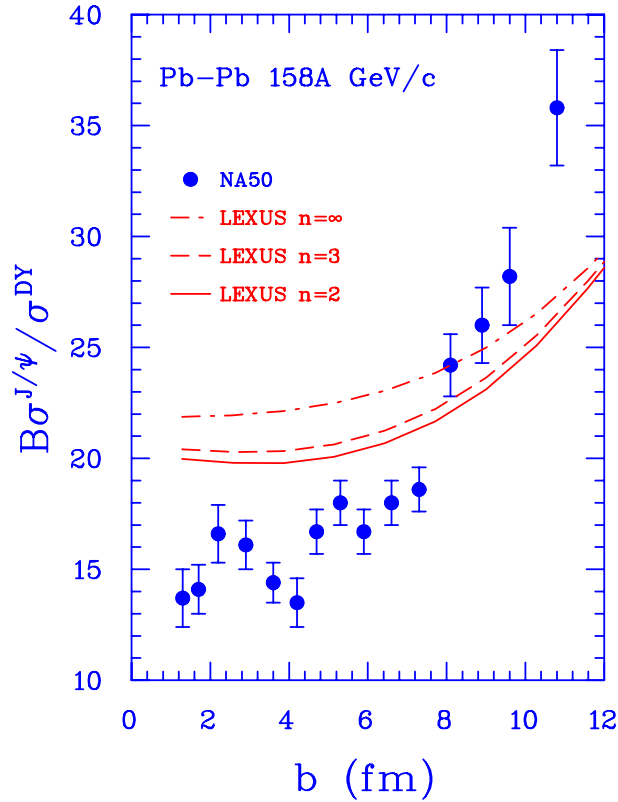}
\end{center}
\caption{Same as Fig. \ref{fig1}, except that this is for the
heavier system. The data is from Ref. \protect\cite{na50_ratio}.}
\label{fig2}
\end{figure}

We now turn to experimental results obtained by the NA50 collaboration 
with Pb projectiles and targets. From Table \ref{table1} we 
see that the measured Drell-Yan cross section 
exceeds our larger ($n$ = 3) value by about one standard deviation. 
The experimental $J/\psi$ value falls within the
predicted range. Plotting the 
$J/\psi$-to-Drell-Yan ratio 
against the impact parameters determined by the experimental collaboration
\cite{na50_ratio} one obtains Fig. \ref{fig2}. Application of this 
model with its parameters determined solely from $pA$ physics does not yield
a satisfactory representation of this experimental data. The latter 
is not reproduced in
trend nor is it in magnitude. Note, however, that the poor quality of this fit is
entirely comparable with those obtained with other  hadronic approaches
\cite{klu99}. Also shown in this figure is the effect of the coherence time
on this ratio. It appears that this effect is not as spectacular here as it
partially cancels in the numerator and denominator. The flattening and
slight increase of the ratio, as one goes to smaller impact parameters, 
can be attributed
to the $J/\psi$ cross section growing slightly faster than the Drell-Yan.
Here also,  the calculated absolute cross sections with no energy loss 
far exceed the experimental values. 

We also considered 
possible nuclear structure effects on the ratio shown in Fig. \ref{fig2}. 
It is known that
parton distribution functions that have a flavor-asymmetric  Dirac sea,
like the one we use in this work,  will
yield different Drell-Yan cross sections depending on whether one has p +
p, p + n, n + n or n + p collisions. In our treatment the
isospin content of the nucleus is assumed to be uniformly distributed
according to the overall charge of the colliding partners. 
Experimentally, Pb is known to have a neutron skin \cite{neutskin}
of  0.19 $\pm$ 0.09 fm. This value is in fact too small to have an effect on
the calculations shown here. Finally, it seems useful to point out 
that in Fig. \ref{fig2} the
Pb data does not seem to converge to the vacuum ratio as one moves towards
more peripheral collisions, unlike the measurements of S-induced reactions. 

We have investigated nucleus-nucleus collisions with
a model that incorporates the coherence time associated with the emission of
soft quanta in hadronic interactions. This approach translates into 
lost energy for the formation of hard radiation, such as high-mass Drell-Yan
pairs  and
$J/\psi$. We have obtained results in quantitative agreement with
experimental data for the reaction S on U at 200A GeV/c. The ratio of
$J/\psi$ to Drell-Yan cross sections as a function of collision centrality,  as
well as the total absolute cross sections are reproduced by our model.
Therefore, we can understand Drell-Yan and $J/\psi$ formation in $pA$ and
S+U collisions in terms of the same physics. 
This model fails to reproduce measurements done in connection with the 
heavier Pb+Pb system.

Several points still need to be clarified.  It will be very instructive
to repeat this analysis in partonic variables including nuclear shadowing
\cite{shadow}.  
A systematic exploration of the freedom allowed by the most recent 
high-precision pA measurements \cite{e866} is called for and 
is underway.  Nevertheless,
if the Pb+Pb data stand the test of time, it does not seem possible to
escape the conclusion that $J/\psi$ suppression is caused by high energy density.
Whether it is due to absorption on hadronic co-movers \cite{comov} or quark-gluon
plasma remains an open and exciting question.

\centerline{\bf ACKNOWLEDGEMENTS}

This work was supported in part by the Natural Sciences and Engineering
Research Council of Canada, in part by the Fonds FCAR of the Quebec
Government, in part by the Director, Office of Energy Research, Office of
High Energy and Nuclear Physics, Division of Nuclear Physics, and by the
Office of Basic Energy Sciences, Division of Nuclear Sciences of the U.S.
Department of Energy under contract DE-AC03-76SF00098, and grant 
DE-FG02-87ER40328.

\newpage
\begin{table}[htb]
\caption{The S+U processes are measured by NA38 \protect\cite{na38_2} 
in collisions at 200A GeV/c and
the Pb+Pb processes are measured by NA50 \protect\cite{na50} in
collisions at 158A GeV/c. The entries in columns labeled with different 
values of $n$
represent calculated values. The $J/\psi$ cross sections are obtained using a
$J/\psi$-nucleon absorption cross section of $\sigma_{\rm abs}$ = 3.6 mb. }
\label{table1}
\newcommand{\m}{\hphantom{$-$}}
\newcommand{\cc}[1]{\multicolumn{1}{c}{#1}}
\renewcommand{\tabcolsep}{1pc} 
\renewcommand{\arraystretch}{1.2} 
\begin{tabular}{@{}llllll}
\ & Process & $\sigma_{\rm expt.}^{\rm tot}$ & \cc{n=$\infty$} & \cc{n=3} 
& \cc{n=2} \\
\hline
S+U: &  & & & &  \\
\ & Drell-Yan & 310 $\pm$ 10 $\pm$ 25 nb  & \m449 nb & \m328 nb& \m272 nb     \\
\ & $J/\psi$ & 7.78 $\pm$ 0.04 $\pm$ 0.62 $\mu$b  & \m12.2 $\mu$b & \m8.38
$\mu$b   
& \m6.83 $\mu$b \\
Pb+Pb: & & & & & \\
\ & Drell-Yan & 1.49$\pm$ 0.01 $\pm$ 0.11 $\mu$b  & \m1.97 $\mu$b & \m1.22
$\mu$b & \m0.95 $\mu$b \\
\ & $J/\psi$ & 21.9$\pm$ 0.02 $\pm$ 1.6 $\mu$b  & \m44.5 $\mu$b & \m26.1
$\mu$b & \m19.8 $\mu$b \\
\end{tabular}
\end{table}

\end{document}